%% file: Capri2010.tex
\documentclass[fleqn,twoside]{article}
\usepackage{espcrc2}

\usepackage{graphicx}
\usepackage[figuresright]{rotating}

\newcommand{\AmS}{{\protect\the\textfont2
  A\kern-.1667em\lower.5ex\hbox{M}\kern-.125emS}}

\title{B physics at CDF --  the Beauty of hadron collisions}

\author{D.~Tonelli\address{Fermilab, \\ P.O. Box 500,  Batavia, IL, 60510, USA}%
        \thanks{for the CDF Collaboration}}
\input{macro}

\newcommand{\betas}{\ensuremath{\beta_{s}^{J/\psi\phi}}}   	% the actual phase that is measured in psiphi Bs decays
\newcommand{\betasSM}{\ensuremath{\beta_{s}^{\mathrm{SM}}}}	% the SM value of the Bspsiphi phase
\newcommand{\betasNP}{\ensuremath{\beta_{s}^{\mathrm{NP}}}}	% the NP contribution to the Bspisphi phase

\newcommand{\phis}{\ensuremath{\phi_{s}^{J/\psi\phi}}}   	% quantities in terms of phi (inteas of beta)
\newcommand{\phisSM}{\ensuremath{\phi_{s}^{\mathrm{SM}}}}

\newcommand{\DGSM}{\ensuremath{\Delta\Gamma^{\mathrm{SM}}}}
       
\begin{document}

\begin{abstract}
The CDF experiment at the Tevatron \pap\ collider established that extensive and detailed exploration of the $b$--quark dynamics is possible in hadron collisions, with results competitive and supplementary to those from $e^+e^-$ colliders. This provides an unique, rich, and highly rewarding program that is currently reaching full maturity. I report a few recent world-leading results on rare decays, \CP-violation in \bs\ mixing, and $b\to s$ penguin decays.
\vspace{1pc}
\end{abstract}

% typeset front matter (including abstract)
\maketitle

\section{Introduction}

Precise results from the successful $B$--factories experiments disfavor large contributions from non standard model (SM) physics in tree-dominated bottom meson decays. Agreement with the SM within theory uncertainties is also manifest in higher-order processes, such as $K^0$--$\overline{K}^0$ or \bd--\abd\ flavor mixing. The emerging picture confirms the Cabibbo-Kobayashi-Maskawa (CKM) \textit{ansatz}  as the leading pattern of flavor dynamics.  Non-SM contributions, if any,  are small corrections, or appear well beyond the TeV scale (and the LHC reach), or have an unnatural, highly fine-tuned flavor structure that escaped all experimental tests to date. The last chances to avoid such a disappointing \emph{impasse} include the physics of bottom-strange mesons, still fairly unexplored, along with a few rare \bd\ decays, not fully probed at the $B$-factories because of limited event statistics. \par The CDF experiment at the Tevatron is currently leading the exploration of this physics,  owing to \CP-symmetric initial states in $\sqrt{s}=1.96$ TeV \pap\ collisions, large event samples collected by a well-understood detector, and mature analysis techniques. 
CDF is a multipurpose magnetic spectrometer surrounded by $4\pi$ calorimeters
and muon detectors. Most relevant for $B$ physics are the tracking, particle-identification (PID) and muon detectors, and the trigger system.  
Six layers of double-readout silicon microstrip detectors  between 2.5 and 22 cm from the beam,  and a single-readout layer at 1.5 cm radius,  provide precise vertex reconstruction, with approximately 15 (70)~\mum\  resolution in the azimuthal (longitudinal) direction.  A drift chamber 
provides 96 samplings of three-dimensional charged-particles trajectories  between 40 and 140 cm radii in $|\eta|<1$,  for a transverse momentum 
resolution of   $\sigma_{p_{T}}/p_{T}^2 = 0.1\%/$(\pgev). Specific ionization measurements in the chamber 
 allow 1.5$\sigma$ separation between charged kaons and pions, approximately  constant at momenta larger than 2~\pgev. A comparable identification is achieved at lower momenta by an array of scintillator bars at 140 cm radius, which  measure the time-of flight.  
 Muons with $\pt>1.5 (2.2)$ GeV/c are detected by planar  drift chambers at $|\eta|<0.6$ ($0.6<|\eta|<1.0$). \par Low-\pt\  dimuon triggers  select  $J/\psi$, rare $B$, and bottomonia decays.  They collected approximately 40 millions  $J/\psi$ decays in 5~\lumifb\ of data,  used to reconstruct roughly 6,000 \bsjpsiphi,  20,000 $B^0\to J/\psi K^*(892)^0$, and 52,000 $B^+\to \jpsi K^+$ decays. A trigger on charged particles displaced from the primary vertex collects hadronic heavy-flavor decays. It relies on dedicated custom electronics to reconstruct tracks in the silicon with offline-like (48~\mum) impact parameter resolution, within 20 $\mu$s of the trigger latency. This yielded approximately 50 millions  $D^0 \to K^-\pi^+$,  13,000 $\bs \to D_s^-(\pi^+\pi^-)\pi^+$, and 12,000 $B^0 \to K^+\pi^-$ decays in 5~\lumifb\ of data. \par CDF has currently collected 8~\lumifb\ of physics-quality data. The sample size will reach 10~\lumifb in October 2011. Additional 6~\lumifb\ will be collected if the proposed three-year extension will be funded.\par  In the following I report some recent, world-leading results,  selected among those more sensitive to the presence of non-SM particles or couplings. Branching fractions indicate \CP-averages, $K^{*0}$ is shorthand for the $K^*(892)^0$ meson, and charge-conjugate decays are implied everywhere.

\section{Polarization in $\bs\to\phi\phi$ decays}

The $\bs\to\phi\phi$ decay proceeds through a penguin-dominated $b\to s\bar{s}s$ transition and was first detected by CDF a few years back.
We recently improved the measurement of branching fraction using $295 \pm 20$ events reconstructed in 2.9~\lumifb\ of data collected by the displaced-track trigger. We obtain $\br(\bs\to \phi\phi)=[2.40 \pm 0.21\stat \pm 0.27\syst \pm 0.82 (\br)] \times 10^{-5}$  \cite{bsphiphi}. The last uncertainty is due to the uncertainty on  $\br(\bsjpsiphi)$,  used as a reference.\par Three polarization amplitudes enrich the phenomenology of the $\bs \to \phi\phi$ decay. They  correspond to the allowed values of orbital angular momentum ($\ell = 0$, 1, or 2) in the decay of a pseudo-scalar into two vector particles ($B\to VV$).    A determination of  these amplitudes may contribute useful insight into the puzzling picture of $B \to VV$ polarizations,  where experimental data in $B^{0(+)} \to \phi K^{*(+)}$ decays disfavor first-order predictions of small transverse component ($f_T  = \mathcal{O}(m^2_V/m^2_B)\approx 4\%$, where $m$ are the masses), possibly suggesting contributions from non-SM amplitudes \cita{hfag}. % \emph{Ad hoc} solutions may accommodate the discrepancy through low-energy QCD corrections. However, they are either model-dependent or non-conclusive, and the non-SM option remains valid. Knowledge of the polarization of $\bs\to\phi\phi$ is deemed useful to discriminate among models. 
\par We report the first  measurement of $\bs\to\phi\phi$ polarization, using the sample employed for the branching fraction measurement  \cita{bsphiphi-new}. We fit  the mass of the four kaons and their angular distributions (in helicity basis).  The angular acceptance,  extracted from simulation, is validated by measuring the polarization of \bsjpsiphi\ decays to be consistent with results from an independent sample \cite{sin2betas}. The small bias due to decay-length requirements in the trigger is included in the systematic uncertainties.  In analogy with measurements of similar $b\to s$ penguin decays, the results are  at odds with naive theory predictions. We measure $|A_0|^2 = 0.348 \pm 0.041\stat \pm 0.021\syst$, $|A_{||}|^2 = 0.287 \pm 0.043\stat \pm 0.011\syst$, $|A_{\perp}|^2 = 0.365 \pm 0.044\stat \pm 0.027\syst$ for the polarization amplitudes, and $\cos(\delta_{||}) = -0.91 ^{+0.15}_{-0.13}\stat\pm 0.09\syst$ for the phase difference between the $A_{||}$ and $A_{0}$ amplitudes. Further extensions of this analysis to larger samples will provide information on the decay-width difference $\Delta\Gamma$  in \bs\ mesons.

\section{Rare $B \to\mu^+\mu^-$ and  $B\to h\mu^+\mu^-$ decays}
%\section{$A_{\mathit{FB}}(\bd\to K^{*0}\mu^+\mu^-)$ in hadron collisions}
Decays mediated by flavor changing neutral currents,  such as $B^0_{(s)} \to \mu^+\mu^-$ or  $B\to h\mu^+\mu^-$ are highly suppressed in the SM because they occur only through higher order loop diagrams. Their phenomenology provide enhanced sensitivity to a broad class of non-SM contributions.\par
The $B^0_{(s)} \to \mu^+\mu^-$  rate is proportional to the CKM matrix element $|V_{td}|^2 (|V_{ts}|^2)$, and is further suppressed by helicity factors. The SM expectations for these branching fractions are $\mathcal{O}(10^{-9})$,  ten times smaller than the current experimental sensitivity.  An observation of these decays at the Tevatron would unambiguously indicate physics beyond the SM. Or,  even improved exclusion-limits strongly constrain the available space of parameters of several SUSY models.\par
\begin{figure}
\centering
\includegraphics[height=5cm,angle=-90]{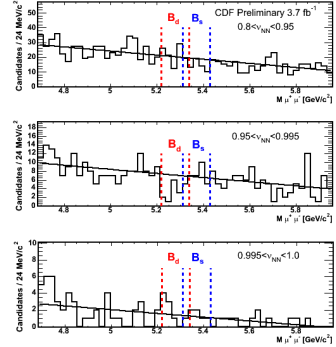}
\caption{\label{fig:bsmass} Distribution of $\mu^+\mu^-$ mass in three independent ranges of the NN classifier.}
\end{figure}

 The latest CDF search for  $B^0_{(s)}\to \mu^+\mu^-$ decays uses  3.7~ \lumifb\ of data collected by the dimuon trigger with  $\pt(\mu)>2~\pgev$ \cite{bsmumu}.  A loose preselection based on opposite-charge dimuon transverse momentum and muon identification criteria (quality of matching with track, energy deposit in the calorimeter, specific ionization energy loss), rejects combinatoric background and charmless $B$ decays. Further rejection is achieved by selecting on the decay-length significance ($\lambda/\sigma_\lambda$) against prompt background and on the isolation of the \bnmeson\ candidate, to exploit the harder fragmentation of \bgmesons\ with respect to light-quark background.  In addition,  we require the candidate to point back to the primary vertex to further reduce combinatoric background and  partially reconstructed \bhadron\ decays. The resulting event sample contains about 55,000 candidates, mostly coming from combinatoric background.  To further enhance purity we use an artificial neural network  classifier (NN) that combines information from the above observables into a single scalar discriminating quantity. Signal distributions are modeled from detailed simulation; backgrounds from mass-sidebands in data.  The signal rate is normalized to 20,000 $\bu \rightarrow \jpsi(\to\mu^+\mu^-) K^+ $ 
reconstructed in the same sample. The ratio of trigger acceptances  between signal and normalization modes ($\simeq 25\%$) is derived from simulation, the relative trigger efficiencies ($\simeq 1$) are extracted  from unbiased data and the relative offline-selection efficiency ($\simeq 80\%$) is determined from simulation and data. The expected average number of background events in the search region is obtained by extrapolating events from the mass-sidebands.  The validity of this extrapolation is checked by comparing predicted and observed background yields in several independent control samples including like-sign dimuons, opposite-sign dimuons with negative decay-length, and opposite-sign dimuons with one muon failing the muon-quality requirements. Small contributions of punch-through hadrons from \bhh\ decays are included in the estimate of total background. We optimize the limit by combining three independent ranges for the NN discriminator (with efficiencies ranging from 44 to 12\%) to obtain the \emph{a priori} best expected 90\% C.L.\ upper limit on $\br(B^0_{(s)}\to \mu^+\mu^-)$ (see \fig{bsmass}).  The resulting 90 (95)\% CL upper-limits are
\begin{eqnarray}
\br(B^0_s \to \mu^+\mu^-) < 3.6 (4.3)\times 10^{-8},\\
\noindent
\br(B^0 \to \mu^+\mu^-) < 6.0 (7.6)\times 10^{-9}.
\end{eqnarray}
These results are the most stringent currently available and reduce  significantly the allowed parameter space for a broad range of SUSY models. An update of this analysis with approximately doubled sample size will further improve them soon. 

We also updated the analysis of $B^0\to K^{*0}(\to K^+\pi^-)\mu^+\mu^-$, $B^+\to K^+\mu^+\mu^-$, 
and $B^0_s\to\phi(\to K^+K^-)\mu^+\mu^-$ decays to 4.4~\lumifb of data \cite{afb}.  These are suppressed in the SM ($\br\approx 10^{-6}$), with amplitudes dominated by penguin and box $b\to s$ transitions.  Despite the presence of final-state hadrons, accurate predictions greatly sensitive to non-SM contributions are possible for relative quantities based on angular-distributions of final state particles. The $B^+$ and $B^0$ decays have been previously studied, while the $B^0_s$ channel has not been observed yet. We use a large sample of high-purity dimuon candidates combined in a common vertex with one or two charged particles, whose masses are assigned as appropriate for each final state. In the \bd\ decay, the mass hypothesis yielding the $K\pi$ mass closer to the known $K^{*0}$ mass is chosen, resulting in the proper assignment  about 92\% of the times.  A NN that uses information from kinematics and particle-identification greatly improves signal-to-background discrimination. It is trained on simulated signals and mass-sideband data. The simulation is tuned to reproduce accurately the data using  the corresponding resonant channels ($B \to J/\psi h$). The NN is optimized for maximum expected statistical resolution on the branching ratio and asymmetry measurements.  \par
\begin{figure}
\includegraphics[height=4.cm,angle=0]{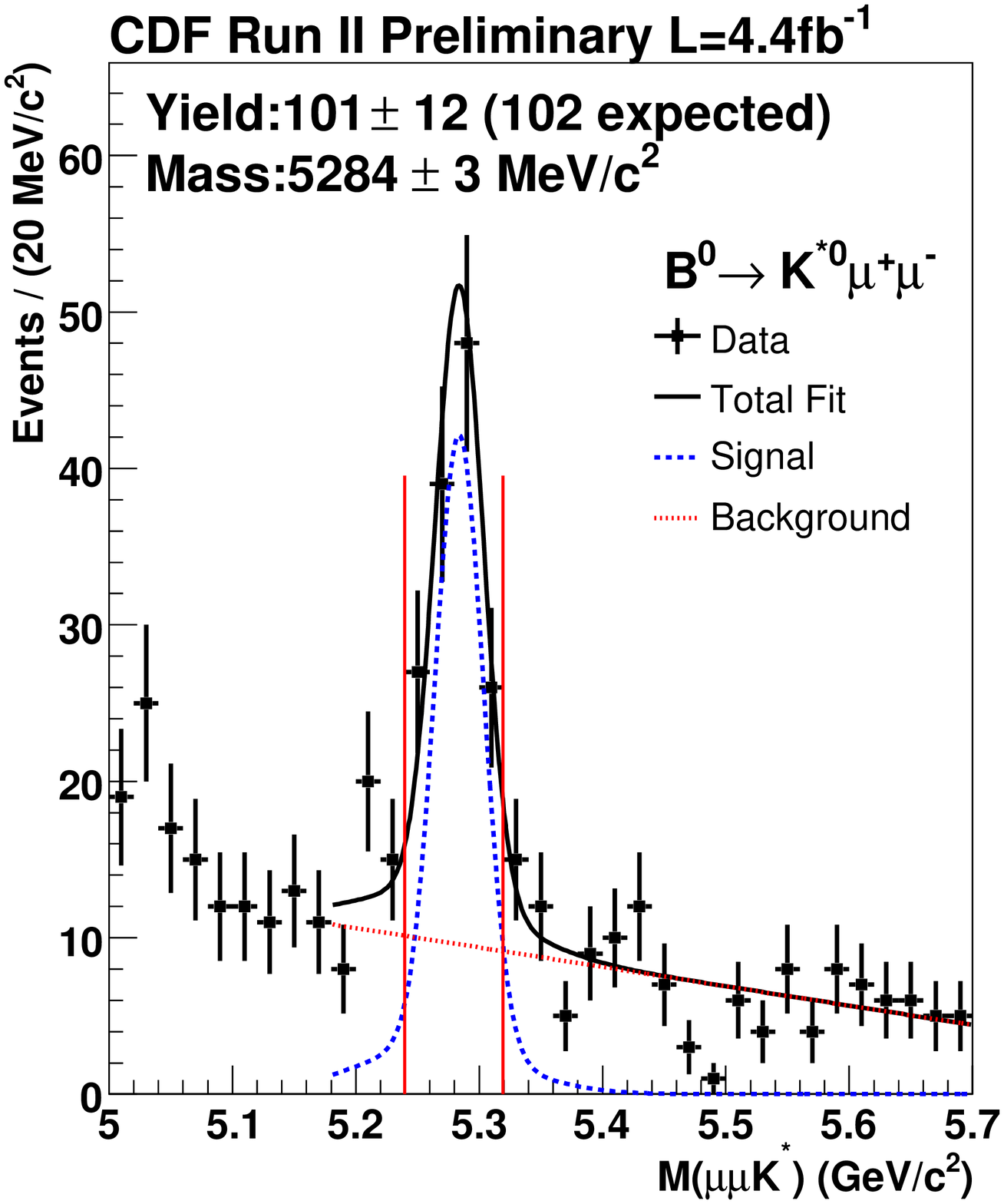}\hspace{.5cm}
\includegraphics[height=4.cm,angle=0]{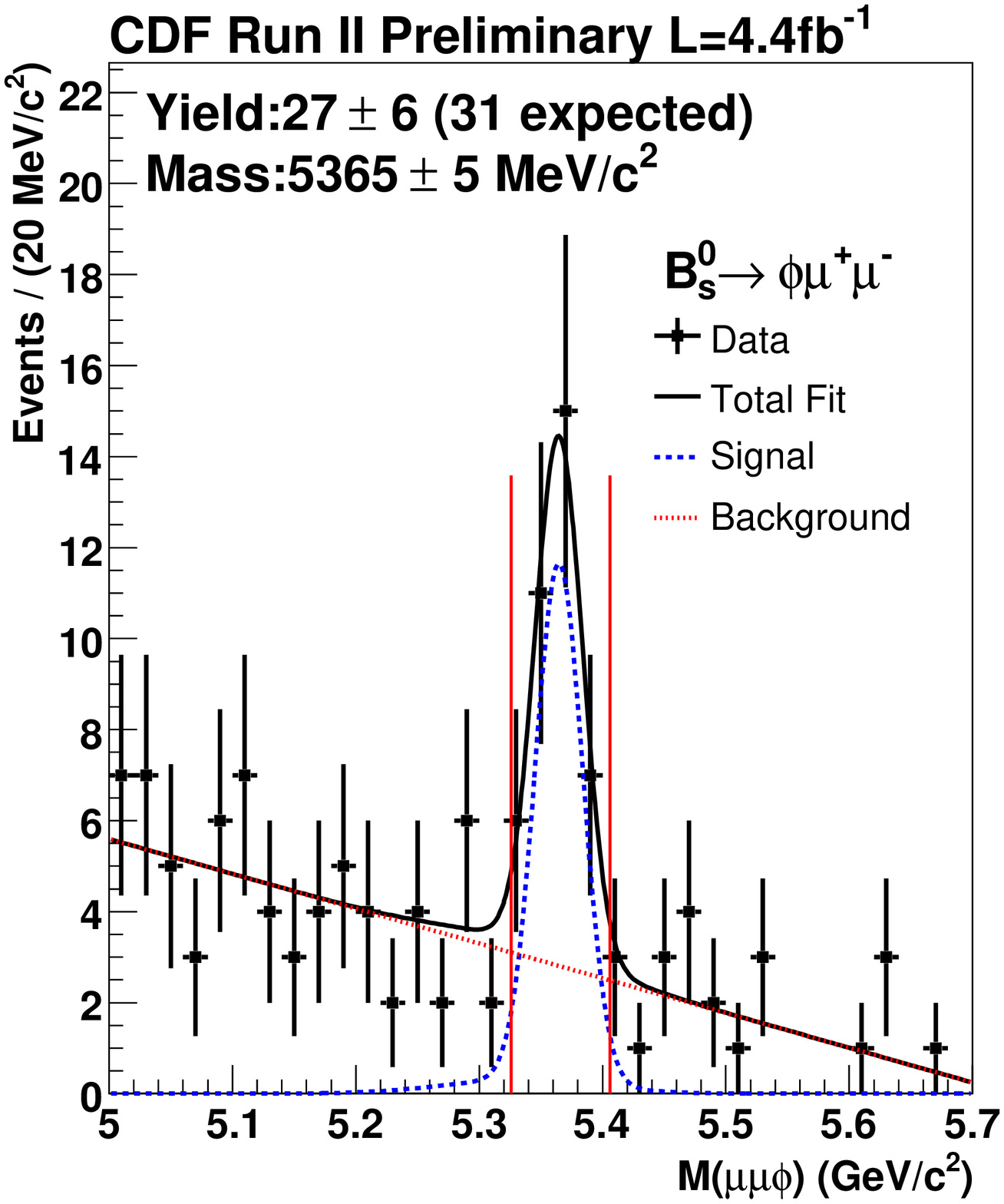}
\caption{\label{fig:bsphimumu} Distribution of $K^+\pi^-\mu^+\mu^-$ mass (left) and $K^+K^-\mu^+\mu^-$ (right).}
\end{figure}

\begin{figure}
\centering
\includegraphics[height=4.4cm,angle=0]{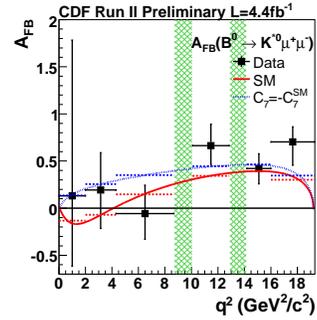}
\caption{\label{fig:afb} Forward-backward asymmetry in $\bd\to K^{*0}\mu^+\mu^-$ as a function of dimuon mass.}
\end{figure}

 Prominent signals of $120\pm16$ $B^+\to K^+\mu^+\mu^-$ and $101\pm12$ $B^0\to K^{*0}\mu^+\mu^-$ events   (\fig{bsphimumu}, left) are observed. The absolute branching fractions, measured using the resonant decays as a reference,  are $[0.38 \pm 0.05\stat \pm 0.03\syst] \times 10^{-6}$ and $[1.06 \pm 0.14\stat \pm 0.09\syst] \times 10^{-6}$, respectively, consistent and competitive with previous determinations \cita{hfag}. In addition, $27\pm 6$ $B^0_s\to\phi\mu^+\mu^-$ events are reconstructed, corresponding to the first observation of this decay, with a significance in excess of $6\sigma$.  The branching ratio, $(1.44\pm 0.33\stat \pm 0.46\syst)\times 10^{-5}$, is consistent with theory predictions, and corresponds to the rarest \bs\ decay ever observed to date. We use the \bu\ and \bd\ signals for the first measurement in hadron collisions of branching ratios, muon forward-backward asymmetry ($A_{\mathit{FB}}$),  and $K^{*0}$ longitudinal polarization, as a function of the dimuon mass. The asymmetry is greatly sensitive to non-SM particles and is determined from a fit to the $\cos(\theta_\mu)$ distributions, $\theta_\mu$ being the helicity angle between the $\mu^+$ ($\mu^-$) and the opposite of the $B$ ($\bar{B}$) direction in the dimuon rest frame. 
%The polarization uses $\cos(\theta_K)$, $\theta_K$ being  the angle between the kaon direction and the direction opposite to the $B^0$ meson in the $K^{*0}$ rest frame. 
Angular acceptances are determined from simulation. Figure~\ref{fig:afb} shows the asymmetry as a function of dimuon mass. Integrated in the 1--6 GeV$^2$/$c^4$ range, where theory predictions are most reliable, it equals  $A_{\mathit{FB}}(1 < q^2 < 6) = 0.43 ^{+0.36}_{-0.37} \stat \pm 0.06\syst$, consistent with Belle and Babar determinations \cita{hfag}. CDF plans to achieve world-leading results by summer 2011 with the anticipated 2--3 factors increase in statistics, due to additional data, triggers, and reconstructed final states. 
%Rare, all-leptonic, flavor changing neutral current decays as $B^0_{(s)}\to \mu^+\mu^-$ are powerful probes for the presence of non-SM physics,  complementary to many direct searches. In the SM they are suppressed below the sensitivity of Tevatron experiments (\eg\ $\br(B^0_s \to \mu^+\mu^-)\approx (3.6 \pm 0.3) \times 10^{-9}$) by CKM and helicity factors but a broad class of SM extensions predict enhancements up to three orders of magnitude for their rates. 
%In minimal supersymmetric (SUSY) extensions of the SM, for instance,  additional processes involving 
%virtual SUSY particles imply $\br(B^0_{(s)}\to \mu^+\mu^-)\propto \tan^6(\beta)$; $\tan(\beta)$ is the ratio of vacuum expectation  values of the two neutral \CP-even Higgs fields; hence large enhancements of the decay rate are expected in those SUSY models, like minimal SO(10) \cite{bmumu_3,bmumu_4,bmumu_5,bmumu_6}, that favor higher values of $\tan(\beta)$ \cite{bmumu_7,bmumu_8,bmumu_9}.
% On the other hand, R-parity violating SUSY models \cite{bmumu_9} may enhance rates even at lower values of $\tan(\beta)$. 

\section{Measurement of the \bs\ mixing phase}
Non-SM contributions have not yet been excluded in \bs-\abs\ mixing. Their magnitude is constrained to be small by the precise determination of the  frequency \cita{mixing}. However, knowledge of only the frequency leaves possible non-SM contributions to the (\CP-violating) mixing phase unconstrained. The time evolution of  flavor-tagged \bsjpsiphi\ decays allow a determination of this phase largely free from theory uncertainties. These decays probe the phase-difference between the mixing and the $\bar{b}\to \bar{c}c\bar{s}$ quark-level transition, $\betas = \betasSM+\betasNP$, which equals $\betasSM=\arg(-V_{ts}V_{tb}^{*}/V_{cs}V_{cb}^{*}) \approx 0.02$ in the SM  and is extremely sensitive to non-SM physics in the mixing. A non-SM contribution  ($\betasNP$)  would also enter $\phis = \phisSM - 2\betasNP$, which is the phase difference between mixing and decay into final states common to \bs\ and \abs, and is also tiny in the SM: $\phisSM = \arg(-M_{12}/\Gamma_{12}) \approx 0.004$. Because the SM values for \betas\ and \phis\ cannot be resolved with the precision of current experiments, the following approximation is used: $\phis \approx -2\betasNP \approx -2\betas$, which holds in case of sizable non-SM contributions. Note that the phase \phis\ also modifies the decay-width difference between light and heavy states, $\Delta\Gamma=\Gamma_L-\Gamma_H=2|\Gamma_{12}|\cos(\phis)$, which enters in the \bsjpsiphi\ amplitude and equals $\DGSM \approx 2|\Gamma_{12}| = 0.086 \pm 0.025$ ps$^{-1}$ in the SM \cita{nierste} . \par We updated the measurement of the time-evolution of flavor-tagged $\bs \to \jpsi(\to\mu^+\mu^-) \phi(\to K^+K^-)$ decays to a sample of 5.2~\lumifb\ collected by the dimuon trigger \cite{sin2betas}. Improvements over the previous version of the analysis include (a) a doubled event sample along with a newly optimized selection  (b) a fully data-driven recalibration of the flavor-tagging algorithms (c) inclusion of possible non-$\phi$ scalar $K^+K^-$  contributions as $\bs \to J/\psi f_0(980)$.\par 
\begin{figure}[h]
\includegraphics[height=3.cm,angle=0]{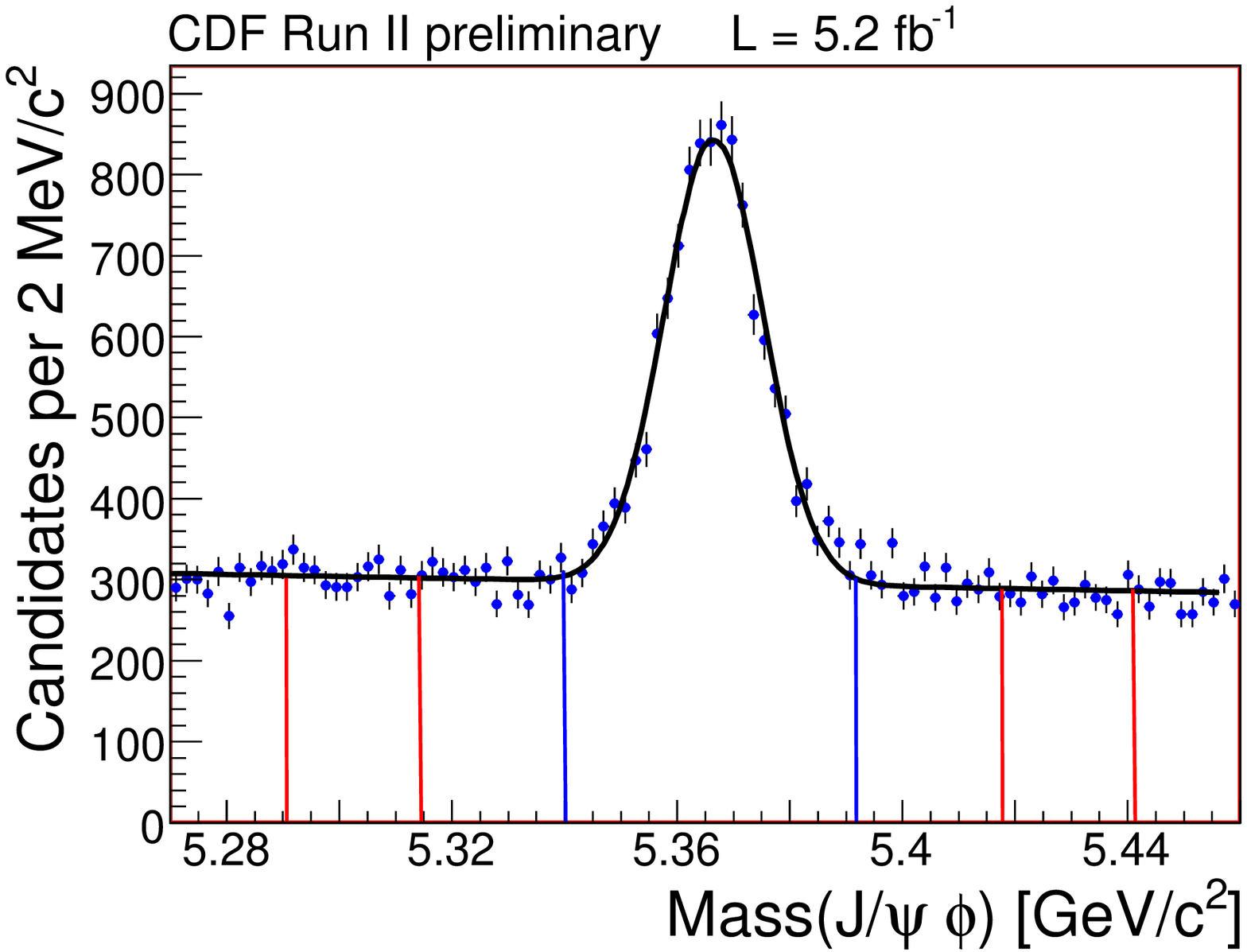}
\includegraphics[height=3.cm,angle=0]{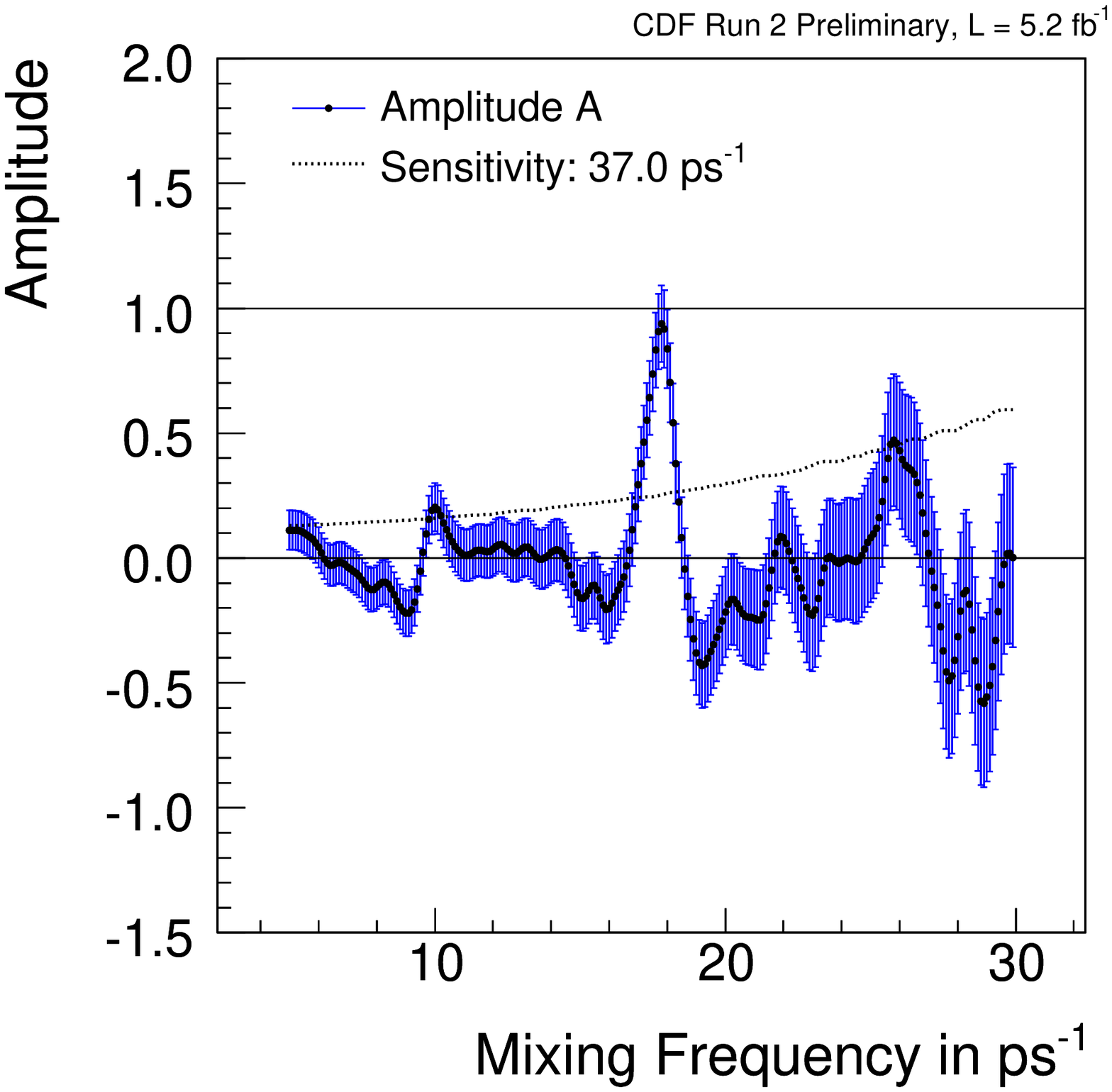}
\caption{\label{fig:sin2betas} Distribution of $\jpsi K^+K^-$ mass after the optimized selection.}
\end{figure}
%\begin{figure}[h]
%\includegraphics[height=3.cm,angle=0]{BsMass.eps}
%\includegraphics[height=3.cm,angle=0]{SSKT.eps}
%\caption{\label{fig:sin2betas} Distribution of $\jpsi K^+K^-$ mass after the optimized selection.}
%\end{figure}
A kinematic fit to a common space-point is applied to the candidate \jpsi\ and two tracks consistent with being kaons originated from a $\phi$ meson decay. A NN trained on simulated data (to identify signal) and \bs\ mass sidebands (for background) exploits kinematic and PID information for an unbiased optimization of the selection: we minimize the average expected statistical uncertainty on the mixing phase extracted from ensembles of statistical trials generated with different input values. Discriminating observables used include kaon-likelihood, from the combination of \dedx\ and TOF information; transverse momenta of the \bs\ and $\phi$ mesons; the $K^+K^-$ mass; and the quality of the vertex fit. The final sample contains approximately 6500 signal events over a comparable background (\fig{sin2betas}, left). The angular distributions of final state particles provide a statistical determination of the \CP-composition of the signal thus improving sensitivity to the phase. Possible contributions from scalar $\jpsi K^+K^-$ or $\jpsi f_0(980)$ final states are included in the angular distributions. %Angles are expressed in the transversity basis, which allows a convenient separation between \CP-odd and \CP-even terms in the equations of the time-evolution.  
The production flavor is inferred using two classes of algorithms. Opposite-side tags (OST) exploit flavor conservation in strong interaction. Being $b$--quarks predominantly produced in pairs with $\bar{b}$--quarks, the production flavor is inferred from the charge of decay products ($e$, $\mu$, or jet) of the $b$--hadron emitted in the opposite hemisphere of the signal \bs. Same-side tags (SST) exploit flavor conservation in the fragmentation process that leads to the signal \bs\ meson. Because a  $s\bar{s}$ pair will be required to form a meson from the bottom quark, the charge of a kaon kinematically close to the bottom-strange meson is correlated to its flavor.  The performance of flavor taggers is modeled as function of many event properties using simulations; however overall scale factors between the simulation and reality are allowed and extracted from data. The uncertainty in their determination contributes to the final systematic uncertainty of the measurement.  The tagging power, $\epsilon D^2$,  is the product of an efficiency $\epsilon$, the fraction of candidates with a flavor tag, and the square of the dilution $D=1-2w$, where $w$ is the mis-tag probability. We calibrate independently $b$ and $\bar{b}$ OST tags using 52,000  $B^+\to\jpsi K^+$ decays. Predicted and measured dilutions agree with scale factors close to unity. The observed tagging efficiency is $(94.2\pm0.4)\%$  and the average predicted dilution on signal is 
$\sqrt{\langle D^2\rangle}=0.110\pm0.002$, with $1.03\pm0.06$ scale factor. The SST calibration is obtained by repeating the full mixing analysis on 13,000 $\bs\to D^-_s\pi^+(\pi^+\pi^-)$ \cita{mixing}. The resulting frequency, $\Delta m_s = 17.79 \pm 0.07\stat$ ps$^{-1}$,  is fully consistent with published values (fig.~\ref{fig:sin2betas}, right). The scale factor, $A=0.94\pm0.15\stat \pm0.13\syst$ is given by the size of the amplitude at the mixing frequency. This corresponds to a tagging efficiency of $(52.2\pm0.7)\%$ and an average dilution on signal of $\sqrt{\langle D^2\rangle}=0.275\pm0.003$.  Multiple tags, if any, are combined as independent for a total tagging power $\epsilon D^2 \approx 4.5\%$. The proper time of the decay and its resolution are known on a per-candidate basis with an average resolution of approximately 90 fs$^{-1}$. Information on \bs\ candidate mass and its uncertainty, angles between final state particles' trajectories (to extract the \CP-composition), production flavor, and decay length and its resolution are used as observables in a multivariate unbinned likelihood fit of the time evolution that accounts for direct decay amplitude, mixing followed by  the decay, and their interference. Direct \CP-violation is expected small and neglected. The fit determines the phase $\betas$, the decay-width difference $\Delta\Gamma$, and many other ``nuisance'' parameters including the mean \bs\ lifetime  ($2/(\Gamma_L + \Gamma_H)$), the magnitudes of linear polarization amplitudes, the \CP-conserving  phases ($\delta_{\parallel} = \arg(A_{\parallel} A_{0}^{*})$, $\delta_{\perp} = \arg(A_{\perp}A_{0}^{*})$), and others.
The acceptance of the detector is calculated from simulation and found to be consistent with angular distributions of random combinations of four tracks in data; the fit model was validated by measuring lifetime and polarization amplitudes in \bdjpsikstar\ decays to be consistent with $B$--factories measurements \cita{jpsikstar}.  \par For enhanced precision, all parameters but the mixing phase are determined in a fit with phase fixed to zero: $c\tau(\bs) = 458.6\pm 7.6\stat \pm 3.6\syst~\mum$, $\Delta\Gamma = 0.075 \pm 0.035\stat \pm 0.010\syst~ \mbox{ps}^{-1}$, $|A_{||}|^2 = 0.231 \pm 0.014\stat\ \pm 0.015 \syst$,  $|A_{0}|^2 = 0.524 \pm 0.013\stat\ \pm 0.015 \syst$,and $\delta_{\perp} = 2.95 \pm 0.64 \stat \pm 0.07\syst$ rad. These results represent the current best measurements of these quantities from a single experiment. \par 
With phase floating, fits on simulated samples show biased, non-Gaussian distributions of estimates and multiple maxima, because of known likelihood symmetries. We use a frequentist confidence region in the ($\betas, \Delta\Gamma$) plane using a profile-likelihood ratio ordering \cita{ichep},  which is close to optimal for limiting the impact of systematic uncertainties. These are included by randomly sampling a limited number of points in the space of all nuisance parameters. A specific value $(\betas, \Delta\Gamma)$ is excluded only if it can be excluded for any assumed value of the nuisance parameters within $5\sigma$ of their estimate on data.  \par The resulting allowed region is greatly reduced with respect to the previous measurement and is fairly consistent ($0.8\sigma$) with the SM:  the range [0.02, 0.52]$\cup$[1.08, 1.55]  contains the \betas\ phase at the 68\% CL and the range [-$\pi$/2, -1.44] $\cup$[-0.13, 0.68]$\cup$[0.89, $\pi$/2] at the 95\% CL.  \par With the full Run II data sample, we expect to observe a non-SM phase or exclude it for any value of $\betas$ larger than approximately 0.4 rad. 
%\begin{figure}
%\includegraphics[height=3.cm,angle=0]{BsMass.eps}
%\includegraphics[height=3.cm,angle=0]{SSKT.eps}
%\caption{\label{fig:sin2betas} Distribution of $\jpsi K^+K^-$ mass after the optimized selection in 5.2 \lumifb\ of CDF data.}
%\end{figure}
%\begin{figure}
%\includegraphics[height=5cm,angle=0]{SSKT.eps}
%\caption{\label{fig:sskt} \bs mixing amplitude scan with 5.2 \lumifb\ of CDF data.}
%\end{figure}
\begin{figure}
\centering
\includegraphics[height=4.4cm,angle=0]{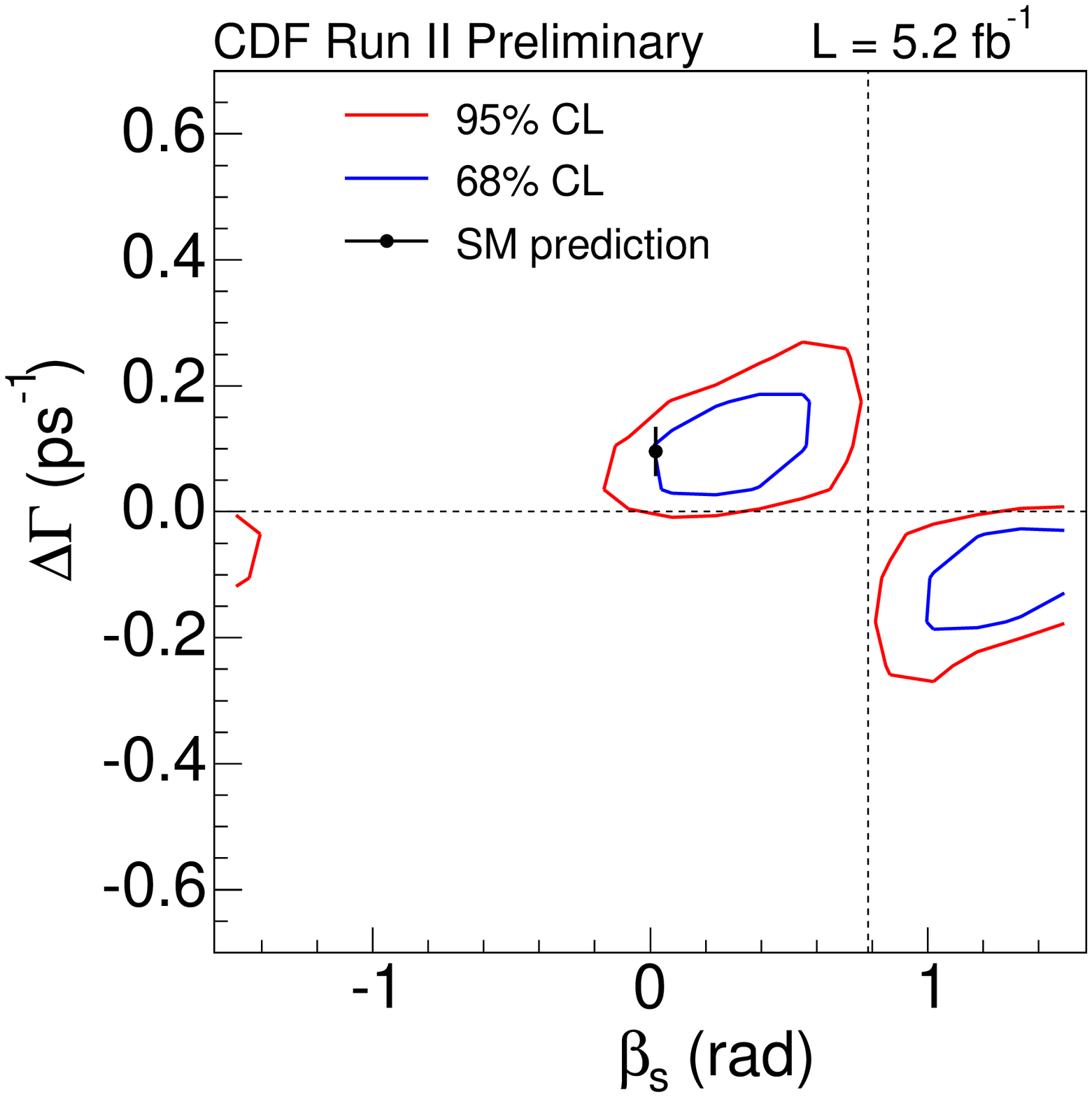}
\caption{\label{fig:contours}Confidence region in the $(\betas,\Delta\Gamma)$ plane.}
\end{figure}

\end{document}

%% file: macro.tex
\def \rightdownarrow
 {\kern.3em
 \rule[.5ex]{.15mm}{2ex}
 {\mbox{$\kern-0.1em{\longrightarrow}$}}      % down-right arrow for cascade decay equations
 }

\def\lessim{\mathrel {\vcenter {\baselineskip 0pt \kern 0pt  % minore/circa uguale
\hbox{$<$} \kern 0pt \hbox{$\sim$} }}}

\def\gessim{\mathrel {\vcenter {\baselineskip 0pt \kern 0pt   % maggiore/circa uguale
\hbox{$>$} \kern 0pt \hbox{$\sim$} }}}

\newcommand{\lumifb}{\mbox{fb$^{-1}$}}				%	fb^-1

\newcommand{\mum}{\mbox{$\mu$m}}				%	um

\newcommand{\br}{\ensuremath{\mathcal{B}}}

\newcommand{\tev}{\ensuremath{\mathrm{Te\kern -0.1em V}}}
\newcommand{\gev}{\ensuremath{\mathrm{Ge\kern -0.1em V}}}	%	GeV
\newcommand{\mev}{\ensuremath{\mathrm{Me\kern -0.1em V}}}	%	MeV
\newcommand{\kev}{\ensuremath{\mathrm{ke\kern -0.1em V}}}	%	keV
\newcommand{\pgev}{\mbox{\gev/$c$}}				%	GeV/c
\newcommand{\stat}{\ensuremath{\mathit{~(stat.)}}}		%	(stat.)
\newcommand{\syst}{\ensuremath{\mathit{~(syst.)}}}		%	(syst.)

\newcommand{\CP}{\ensuremath{\mathsf{CP}}}			%	CP

\newcommand{\pap}{\proton\antiproton}			%	ppbar
\newcommand{\pt}{\ensuremath{p_{\rm{T}}}}			%	pT

\newcommand{\proton}{\ensuremath{\mathrm{p}}}
\newcommand{\antiproton}{\ensuremath{\bar{\rm{p}}}}

\newcommand{\bd}{\ensuremath{B^{0}}}				% 	B-zero
\newcommand{\bs}{\ensuremath{B^{0}_s}}				% 	B (zero) sub-s

\newcommand{\bu}{\ensuremath{B^{+}}}				%	B piu'
\newcommand{\abd}{\ensuremath{\overline{B}^{0}}}		% 	B-zero
\newcommand{\abs}{\ensuremath{\overline{B}^{0}_s}}		% 	B (zero) sub-s
\newcommand{\bhadron}{\mbox{$b$--hadron}}			%	b-hadron
\newcommand{\bgmesons}{\mbox{$b$--mesons}}			% 	generic B

\newcommand{\bn}{\ensuremath{B^{0}_{(s)}}}			% 	neutral B (B-zero and B sub-s)
\newcommand{\bnmeson}{\mbox{$B^0_{(s)}$ meson}}			% 	neutral B (B-zero and B sub-s)
\newcommand{\bhh}{\ensuremath{\bn \to h^{+}h^{'-}}}

\newcommand{\bdjpsikstar}{\ensuremath{\bd \to  \jpsi K^{*0}}}

\newcommand{\bsjpsiphi}{\ensuremath{\bs \to  \jpsi \phi}}

\newcommand{\jpsi}{\ensuremath{J/\psi}}

\newcommand{\fig}[1]{fig.~\ref{fig:#1}}

\newcommand{\cita}[1]{\cite{#1}}

\newcommand{\dedx}{\ensuremath{\mathit{dE/dx}}}

\newcommand{\arxiv}[1]{\textsf{[#1]}}
\def\babar{\mbox{\slshape B\kern-0.1em{\smaller A}\kern-0.1em B\kern-0.1em{\smaller A\kern-0.2em R}}}

%% file: Capri2010.bbl
\begin{thebibliography}{99}
\bibitem{bsphiphi} CDF Collaboration, Note 10064, and  D.~Acosta \emph{et al.} (CDF Collaboration), Phys.\ Rev.\ Lett.\ 95, 031801 (2005).
\bibitem{hfag} Heavy Flavor Averaging Group, arXiv:\arxiv{1010.1589(hep-ex)}.
\bibitem{bsphiphi-new} CDF Collaboration, Note 10120.
\bibitem{sin2betas} CDF Collaboration, Note 10206,  and T.~Aaltonen \emph{et al.} (CDF Collaboration), Phys.\ Rev.\ Lett.\  100, 161802 (2008).
\bibitem{bsmumu} CDF Collaboration, Note 9892, and T.~Aaltonen \emph{et al.} (CDF Collaboration), Phys.\ Rev.\ Lett.\ 100, 101802 (2008).
\bibitem{afb} CDF Collaboration, Note 10047,   and T.~Aaltonen \emph{et al.} (CDF Collaboration), Phys.\ Rev.\ D79, 011104(R) (2009).
\bibitem{mixing} CDF Collaboration, Note 10108, and A.~Abulencia \emph{et al.} (CDF Collaboration), Phys.\ Rev.\ Lett.\  97, 242003 (2006).
\bibitem{nierste} U. Nierste, talk at CKM Workshop 2010.
\bibitem{jpsikstar} CDF Collaboration,  Note 8950. 
\bibitem{ichep} D.~Tonelli (for the CDF Collaboration), arXiv:\arxiv{0810.3229(hep-ex)}.
\end{thebibliography}
